\documentclass[fleqn,12pt,twoside]{article}
\usepackage{espcrc1}

\usepackage{graphicx}
\usepackage[figuresright]{rotating}

\newcommand{\AmS}{{\protect\the\textfont2
  A\kern-.1667em\lower.5ex\hbox{M}\kern-.125emS}}

\title{Soliton propagation in relativistic hydrodynamics}

\author{D.A. Foga\c{c}a\address[MCSD]{Instituto de F\'{\i}sica, Universidade de S\~{a}o Paulo\\
                                  C.P. 66318,  05315-970 S\~{a}o Paulo, SP, Brazil.}
        and
        F.S. Navarra\addressmark[MCSD]
}

\begin{document}

\maketitle

\begin{abstract}
We study the conditions for the formation and propagation of Korteweg-de Vries (KdV) solitons
in nuclear matter. In a previous work we have derived a  KdV equation from Euler
and continuity equations in non-relativistic  hydrodynamics.  In the present contribution
we extend our formalism to  relativistic fluids.  We present results for a given equation of
state, which is based on quantum hadrodynamics (QHD).

\end{abstract}

\section{Introduction}

Long ago \cite{frsw} it was suggested that Korteweg - de Vries solitons might be formed
in the  nuclear medium. In a  previous work \cite{nois}  we have updated the early works
on the subject introducing a realistic equation of state (EOS) for nuclear matter.  We have
found that these solitary waves can indeed  exist in the nuclear medium, provided that
derivative couplings between the nucleon and the vector field are included. These couplings
lead to an
energy density which depends on the  Laplacian of the baryon density. For this class of
equations  of state, which is quite general as pointed out in \cite{furn}, perturbations
on the nuclear density can propagate as a pulse without dissipation.

During the analysis of
several realistic nuclear equations of state, we realized that, very often the speed of sound
$c_s^2$ is in the range $0.15 -0.25$. Compared to the speed of light these values are not large
but not very small either. This suggests that, even for slowly moving nuclear matter,
relativistic effects might be sizeable. This concern justifies the extension of the
formalism presented in \cite{nois}.

\section{Hydrodynamics}

Euler equation and the continuity equation form the basis of hydrodynamics.
In the non-relativistic regime and for a perfect fluid  they are \cite{nois}:
\begin{equation}
{\frac{\partial \vec{v}}{\partial t}} +(\vec{v} \cdot \vec{\nabla}) \vec{v}=
-\bigg({\frac{1}{M}}\bigg) \vec{\nabla} h
\label{eulerentalpia}
\end{equation}
\begin{equation}
{\frac{\partial \rho_B}{\partial t}} + {\vec{\nabla}} \cdot (\rho_B {\vec{v}})=0
\label{contibari}
\end{equation}
where $\rho_B$, $M$, $h$ and $v$ are the baryon density,  the nucleon mass,  the enthalpy per
nucleon and the fluid velocity respectively. In the relativistic case  they are
\cite{elze,wein}:
\begin{equation}
{\frac{\partial {\vec{v}}}{\partial t}}+(\vec{v} \cdot \vec{\nabla})\vec{v}=
-{\frac{1}{(\varepsilon + p)\gamma^{2}}}
\bigg({\vec{\nabla} p +\vec{v} {\frac{\partial p}{\partial t}}}\bigg)
\label{eul}
\end{equation}
\begin{equation}
{\frac{\partial}{\partial t}}(\rho_{B}\gamma)+\vec{\nabla} \cdot (\rho_{B}\gamma \vec{v})=0
\label{con}
\end{equation}
where $\gamma$, $\varepsilon$ and $p$ are the usual Lorentz factor ($\gamma=(1-v^{2})^{-1/2}$),
energy density and  pressure respectively. We have deliberately written the above equations in
a non-covariant way to make the comparison between the non-relativistic and relativistic cases
easier. Using the definition of enthalpy per nucleon \cite{land} for a perfect fluid we find
that   $dp=\rho_{B}dh$. Therefore, using the Gibbs relation at zero temperature
$\varepsilon + p=\mu_{B} \,  \rho_{B}$ we can rewrite (\ref{eul}) as:
\begin{equation}
{\frac{\partial {\vec{v}}}{\partial t}}+(\vec{v} \cdot \vec{\nabla})\vec{v}= \, - \,
{\frac{(1-v^{2})}{\mu_{B}}}
\bigg({\vec{\nabla} h +\vec{v} {\frac{\partial h}{\partial t}}}\bigg)
\label{eulerfinal}
\end{equation}
where $\mu_B$ is the baryochemical potential.
Since the enthalpy per nucleon may also be written as \cite{nois,abu}
\begin{equation}
h={\frac{\partial \varepsilon }{\partial \rho_{B}}}
\label{gradh}
\end{equation}
it becomes clear that the ``force'' on the right hand side of the Euler equations
(\ref{eulerentalpia}) and (\ref{eulerfinal}) will be ultimately determined by the
equation of state, i.e., by the function  $\varepsilon(\rho_{B})$.

\section{KdV equation and the  nuclear equation of state}

Equations  ({\ref{eulerentalpia}}) and (\ref{eulerfinal})  contain the gradient of the
derivative of the energy density. If $\varepsilon$ contains a
Laplacian of $\rho_B$, i.e., ${\mathcal{\varepsilon}}
\propto ... + ... \nabla^{2} \rho_{B} + ...$, then
({\ref{eulerentalpia}}) and (\ref{eulerfinal}) will have a cubic derivative
with respect to the space coordinate which will give rise to the Korteweg-de Vries equation
for the baryon density.
The most popular  relativistic mean field models do not have higher derivative terms and,
even if they have at the start, these terms are usually neglected during the calculations.

In \cite{nois} we have added a new derivative term to  the usual non-linear QHD
\cite{lala}, given by
\begin{equation}
{\mathcal{L_{M}}} \equiv {\frac{g_{v}}{{m_{v}}^{2}}}\bar{\psi}
(\partial_{\nu} \partial^{\nu} V_{\mu})\gamma^{\mu} \psi
\label{lagram}
\end{equation}
where, as usual, the degrees of freedom are
the baryon field $\psi$, the neutral scalar meson field $\phi$
and the neutral vector meson field $V_{\mu}$, with the respective couplings and masses.
The new term is designed to be small in comparison with the main baryon - vector meson
interaction term $g_{v} \bar{\psi} \gamma_{\mu} V^{\mu}  \psi$.
Folowing the standard steps of the mean field formalism we arrive at the following
expression for the energy density:
\begin{eqnarray}
\varepsilon&=&{\frac{{g_{v}}^{2}}{2{m_{v}}^{2}}}\rho_{B}^{2}
+{\frac{{m_{s}}^{2}}{2}}{\bigg[{\frac{(M^{*}-M)}{g_{s}}}\bigg]}^{2}
+{\frac{\eta}{(2\pi)^{3}}}\int_{0}^{k_{F}} d^3{k} ({\vec{k}}^{2}+{M^{*}}^{2})^{1/2}
+{\frac{b}{3g_s^3}}(M^{*}-M)^{3} \nonumber \\
&+&{\frac{c}{4g_{s}^{4}}}(M^{*}-M)^{4} + {\frac{{g_{v}}^{2}}{{m_{v}}^{4}}}\rho_{B}
\nabla^{2}\rho_{B}
\label{epsilonexp}
\end{eqnarray}
where $\eta$ is  the baryon  spin-isospin degeneracy factor,
$M^*$ stands for the nucleon effective mass (given by $M^{*} \equiv M-g_{s}\phi_{0}$)
and the constants $b$, $c$ $g_s$ and $g_v$ are taken from \cite{lala}.

Although Eq.  (\ref{epsilonexp})
was obtained with the help of a specific Lagrangian taken from \cite{lala} and a prototype
Laplacian interaction (\ref{lagram}), the above  form of the energy density
follows quite naturally from an approach based on the density functional theory \cite{fst97},
regardless of the form of the  underlying Lagrangian. Thus KdV solitons are a general
consequence of many-body dynamics.

\section{KdV solitons}

We now repeat the steps developed in \cite{frsw,nois}  and
introduce dimensionless variables for the baryon density and velocity:
\begin{equation}
\hat\rho={\frac{\rho_{B}}{\rho_{0}}} \hspace{0.2cm}, \hspace{0.5cm} \hat v={\frac{v}{c_{s}}}
\label{varschapeu}
\end{equation}

We next  define the ``stretched coordinates''  $\xi$ and $\tau$ as in
\cite{frsw,abu,davidson}:
\begin{equation}
\xi=\sigma^{1/2}{\frac{(x-{c_{s}}t)}{R}}
\hspace{0.2cm}, \hspace{0.5cm}
\tau=\sigma^{3/2}{\frac{{c_{s}}t}{R}}
\label{stret}
\end{equation}
where $R$ is a size scale and $\sigma$ is a small ($0 < \sigma < 1$) expansion parameter
chosen to be \cite{davidson}:
\begin{equation}
{\sigma} = {\frac{\mid u-{c_{s}} \mid}{{c_{s}}}}
\label{sigma}
\end{equation}
where $u$ is the propagation speed of the perturbation in question.
We then expand (\ref{varschapeu})  around  the equilibrium values:
\begin{equation}
\hat\rho=1+\sigma \rho_{1}+ \sigma^{2} \rho_{2}+ \dots
\label{roexp}
\end{equation}
\begin{equation}
\hat v=\sigma v_{1}+ \sigma^{2} v_{2}+ \dots
\label{vexp}
\end{equation}
After the expansion above (\ref{eulerentalpia}),  (\ref{contibari}), (\ref{con}) and
(\ref{eulerfinal})  will contain power series in
$\sigma$   (in practice we go up to $\sigma^2$). Since the coefficients in these series are
independent of each other we get  a set of equations, which, when combined,
lead to  KdV equations for $\rho_{1}$. In the non-relativistic case we have obtained
\cite{nois}:
\begin{equation}
{\frac{\partial {\rho}_{1}}{\partial \tau}}+
3{{\rho}_{1}}{\frac{\partial{\rho}_{1}}{\partial \xi}}
+\bigg({\frac{{g_{v}}^{2}{\rho_{0}}}{2M{c_{s}}^{2}{m_{v}}^{4}R^{2}}}\bigg)
{\frac{\partial^{3}{\rho}_{1}}{\partial \xi^{3}}}=0
\label{KdVpaper}
\end{equation}
with the analytical solitonic solution:
\begin{equation}
{\hat{\rho}_{1}}(x,t)={\frac{(u-{c_{s}})}{{c_{s}}}}
sech^{2}\bigg[
{\frac{{m_{v}}^{2}}{{g_{v}}}}\sqrt{{\frac{(u-{c_{s}}){c_{s}}M}
{2{\rho_{0}}}}}(x-ut) \bigg]
\label{solpaper}
\end{equation}
where ${\hat{\rho}_{1}} \equiv \sigma {\rho_{1}}$ and $u$ is the velocity of propagation
of the perturbation.  The solution above is a bump with width $\lambda$ given by:
\begin{equation}
\lambda={\frac{g_{v}}{{m_{v}}^{2}}} \sqrt{{\frac{2{\rho_{0}}}{(u-{c_{s}}){c_{s}}M}}}
\label{width}
\end{equation}
Now, following the same sequence of steps,  the combination of (\ref{eulerfinal}) and
(\ref{con}) leads to a similar KdV equation for the relativistic case:
\begin{equation}
{\frac{\partial {\rho}_{1}}{\partial \tau}}+
(3-{c_{s}}^{2})
{{{\rho}_{1}}{\frac{\partial{\rho}_{1}}{\partial \xi}}}
+\bigg({\frac{g_v^2 \rho_0}{2 M c_s^2 m_v^4 R^2}}\bigg)
{\frac{\partial^{3}{\rho}_{1}}{\partial \xi^{3}}}=0
\label{KdVmqhdrelat}
\end{equation}
with the solution given by:
\begin{equation}
{\hat{\rho}_{1}}(x,t)={\frac{3(u-{c_{s}})}{{c_{s}}}}(3-{c_{s}}^{2})^{-1}
sech^{2}\bigg[
{\frac{{m_{v}}^{2}}{{g_{v}}}}\sqrt{{\frac{(u-{c_{s}}){c_{s}}M}
{2{\rho_{0}}}}}(x-ut) \bigg]
\label{solmqhdrelat}
\end{equation}
with the condition $\mu_B=M$.

As a consitency check we take the non-relativistic limit, which, in this
case, means taking a small speed of sound  $c^2_s \rightarrow 0$. In this limit
$(3-{c_{s}}^{2})\cong 3$, (\ref{KdVmqhdrelat}) reduces to  (\ref{KdVpaper})
and (\ref{solmqhdrelat}) coincides with   (\ref{solpaper}).

\section{Conclusions}

The existence of KdV solitons in nuclear matter has potential applications in
nuclear physics at intermediate energies \cite{frsw} and also possibly at high
energies. The experimental measurements of jet quenching and related phenomena
performed at RHIC \cite{star} offer an unique opportunity of studying supersonic
motion in hot and dense hadronic matter. With this scenario in mind we gave the
first step in the adaptation of the KdV soliton formalism to the new environment.
We have extended the results of our previous work \cite{nois},
showing that it is possible to obtain the KdV solitons in  relativistic
hydrodynamics with an appropriate  EOS.  Taking the  non-relativistic limit
($c^2_s \rightarrow 0$)  we were able to  recover the previous results.

\end{document}